\documentclass[12pt]{article}

\usepackage{sbc-template}
\usepackage{graphicx}

\usepackage{parallel}
\usepackage{graphicx,url}
\usepackage[utf8x]{inputenc}
\usepackage[portuguese,english]{babel} 
\usepackage[table,xcdraw]{xcolor}
\usepackage[none]{hyphenat} 
\usepackage{multirow}
\usepackage{longtable}
\usepackage{lscape}
\usepackage{hyperref}
\usepackage{amssymb}
\usepackage{comment} 

\usepackage{subfigure}
\usepackage[T1]{fontenc}
\usepackage{fdsymbol}
\usepackage{amssymb}
\usepackage[normalem]{ulem}
\usepackage{xspace}

\newcommand{\citep}[1]{\protect\citeauthoronline{#1}~(\citeyear{#1})\xspace}

\title{BULNER: BUg Localization with word embeddings and NEtwork Regularization}

\author{Jacson Rodrigues Barbosa\inst{1,5}, Ricardo Marcondes Marcacini\inst{3}, Ricardo Britto\inst{4}, \\Frederico Soares\inst{6}, Solange Rezende\inst{1}, Auri M. R. Vincenzi\inst{2}, Márcio E. Delamaro\inst{1}
}

\address{ICMC-Universidade de São Paulo (USP)
\nextinstitute
DC-Universidade Federal de São Carlos (UFSCar)
\nextinstitute
Universidade Federal de Mato Grosso do Sul (UFMS)
\nextinstitute
Ericsson AB / Blekinge Institute of Technology
\nextinstitute
INF-Universidade Federal de Goiás (UFG)
\nextinstitute
Tribunal de Justiça do Estado de Goiás
\email{jacsonrb@usp.br, ricardo.marcacini@ufms.br, ricardo.britto@[ericsson.com,bth.se]},
\email{fasoares@tjgo.jus.br, \{solange,delamaro\}@icmc.usp.br, auri@dufscar.br}
}

\begin{document} 

\maketitle

\begin{abstract}
Bug localization (BL) from the bug report is the strategic activity of the software maintaining process. Because BL is a costly and tedious activity, BL techniques information retrieval-based and machine learning-based could aid software engineers. We propose a method for BUg Localization with word embeddings and Network Regularization (BULNER). The preliminary results suggest that BULNER has better performance than two state-of-the-art methods.
\end{abstract}

\section{Introduction}
\label{intro}
\begin{sloppypar}
Bug localization (BL) from bug reports is an expensive step in the software life cycle because of the manual process of localization. For example, the Mozilla project receives almost 300 bug reports per day, and each one needs a manual triage. Also, often, a bug report (84-93\% of bugs) impacts one or a few files~\cite{Thung2014BITSFBL}. There is a family of bug localization techniques that uses Information Retrieval (IR). IR suggests defective parts of a software system by automatically relating a bug report's vocabulary and associated source code metrics. IR often uses Vector Space Model (VSM) but, due to VSM limitations, recent studies apply distributional semantics of words~\cite{Rahman2018IIBBL}. 

In this paper, we propose BULNER, an IR-based bug localization method, which stands for BUg Localization with word embeddings and Network Regularization. BULNER considers both word embedding features of bug reports and features extracted from project source file metrics. We combined these features in an information network proposed in BULNER. We present a network regularization-based machine learning method that obtains a more appropriate representation model for identifying potential buggy files from bug report texts. Our research answers the following research questions: \textbf{RQ1} - How effective is BULNER? \textbf{RQ2} - What is the contribution of each model? We carried out an experimental evaluation using three well-known real-world datasets. BULNER is competitive with two other state-of-the-art methods, and the experimental results indicate that combining different representation models, such as information networks and the vector-space model, is a promising method.
\end{sloppypar}

\section{Bug Localization Data Model Representation}
\label{sec:background}

Bug localization has been modeled as an information
retrieval task, where the bug report is treated as a query, and sources code files that conform to the system are documents. The goal is to select the files that better match the query based on a defined similarity measure. The effectiveness of the similarity measure depends on the text representation model of bug reports, often based on Bag-of-Words (BoW) and Word Embeddings (WEmb).

\begin{sloppypar}
\textbf{BoW} representation uses terms (e.g., keywords) extracted from texts as features in a vector space model. BoW is a document-term matrix, where each row (vector) represents a document, each column represents a term (word) present in the document collection, and each cell contains a measure (term frequency and inverse document frequency)~\cite{ref:Tan2005}. BoW representation has as main characteristics of the high dimensionality and the high sparsity. However, these characteristics affect the performance of machine learning algorithms negatively. Furthermore, machine learning algorithms are not able to infer relations between terms or relations between documents as they are not established by BoW.

In bug localization context, each source code is defined as weights' vector in BoW, and they use cosine similarity function to identify closely related vector. Zhou~\textit{et~al.} define BugLocator, an IR-based bug localization method based on revised Vector Space Model. From the initial bug report, BugLocator applies textual similarity using similar bugs' information that was fixed before. Then, it ranks all suspicious sources files~\cite{Zhou2012WSTBB}.

\textbf{WEmb} are a mapping table from words to continuous vectors (e.g., vec(``dog'') = [0.8, 0.3, 0.1], vec(``cat'') = [0.7, 0.5, 0.1] , vec(``pasta'') = [0.2, 0.1, 0.7]). In this example, the first parameter of each word represents some kind of animal. We could calculate the semantic similarity between words by cosine similarity and consequently calculate the similarity between sentences or entire documents. To obtain each above vector, we use Skip-gram model proposed in the word2vec method for language modeling~\cite{Mikolov2013DROWAP}. It is a unsupervised method that define meaning each word in its context (e.g., context(``dog'') = [``Pet,'' ``tail,'' ``smell,''],  context(``cat'') = [``pet,'' ``tail,'' ``home'']). Two sets of context words also have common conceptually. According to the distributional hypothesis, we can estimate how close these two words are to each other by comparing with others in same context~\cite{Mikolov2013DROWAP}.
\end{sloppypar}

Ye~\textit{et~al.} use word embedding to train on software documents (API documents, reference documents, and tutorials). They adapt the Skip-gram model and aggregated software documents to estimate semantic similarities between them. Then, from an initial bug report (query document), the bug localization model computes the ranking score for all source code~\cite{Ye2016FWETDS}.

\section{Proposed Method}
\label{sec:proposed}
\begin{figure}[!ht]
    \centering
    \includegraphics[width=0.90\textwidth]{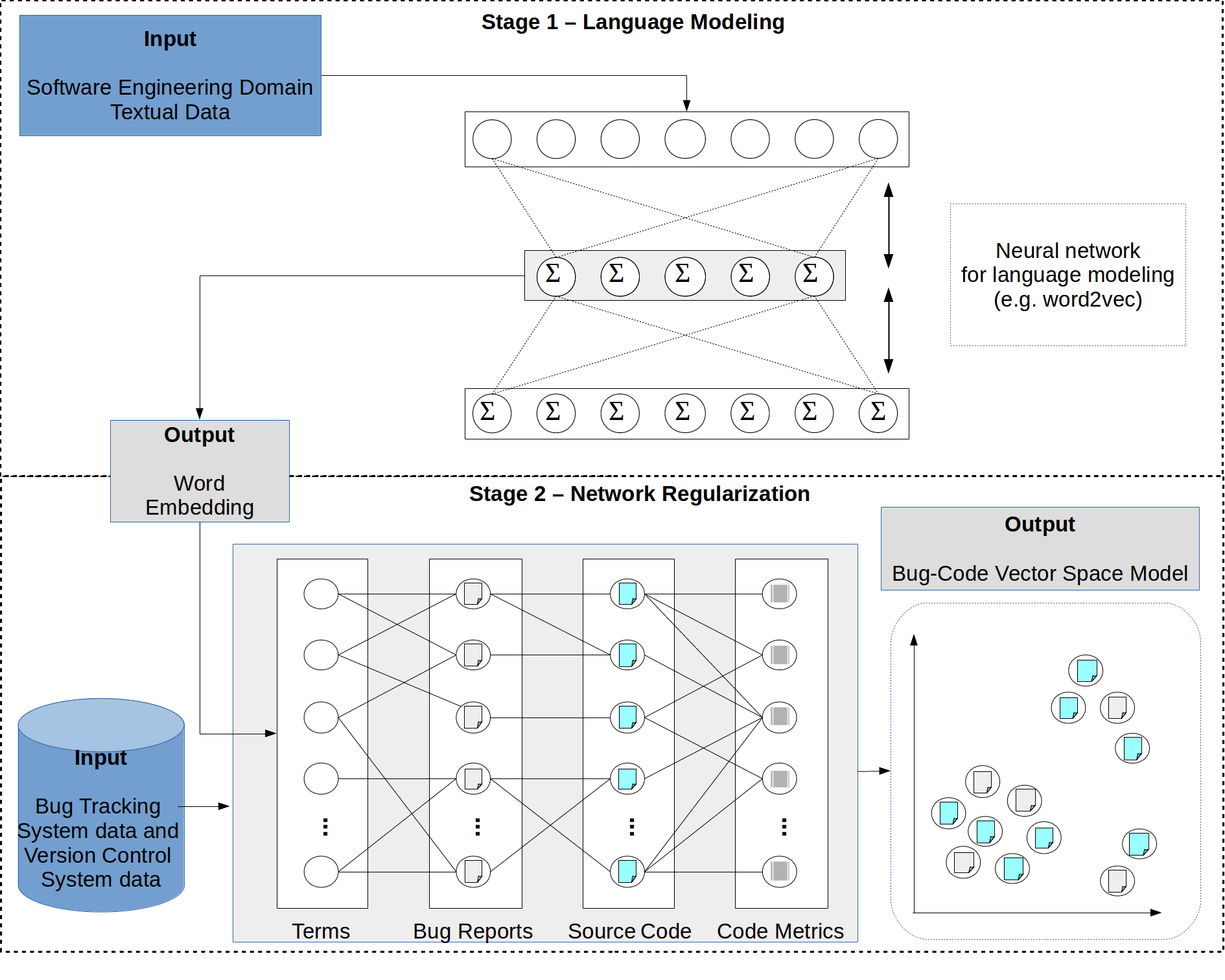}
    \caption{An overview of the BULNER.}
    \label{model}
  \end{figure}

\begin{sloppypar}
In this section, we introduce a new method for Bug Localization, called BULNER. Our method innovates by considering both the semantic content of bug reports through language models and source code content through code metrics. Figure~\ref{model} shows an overview of the BULNER method. The method has two stages: (1) language modeling and (2) network regularization. Given a new bug report, BULNER identifies the most similar bug reports, and the source code files that probably contain the related bug. While the BULNER's first stage enables more accurate computation of the similarity between bug reports, the second performs fine-tuning of the language model by considering relationships between bug reports, source code files, and code metrics.
\end{sloppypar}

Unlike existing methods that calculate the similarity between bug reports through keywords or language models, BULNER learns a new vector space model to directly compares bug reports and source code files. We call this new vector space of Bug-Code Vector Space Model. Moreover, our method allows the inclusion of domain information, such as code metrics, during the learning process of this vector space model.

The first stage of the BULNER method uses a neural network. The primary purpose of this stage is to learn word vector representations from a sizeable textual dataset of the software engineering domain. For example, the term `abort' has the following correlated terms in the language model used in BULNER: `interrupted,' `terminate,' `halt,' `timed-out,' and `exit.' The BULNER method uses the language model based on the skip-gram model trained over 15GB of textual data from Stack Overflow posts, as proposed in~\cite{Efstathiou2018WEFTSED}. The output of the skip-gram model is a representation called word embedding, where each term $t$ contains a representation $\mathbf{Y}(t)$ in the d-dimensional vector space, i.e., $\mathbf{Y}(t) \in \mathbb{R}^d$.

The second stage of the BULNER method uses network regularization to perform the fine-tuning of the model obtained in the first stage. Given a bug reports dataset, we propose a heterogeneous network-based representation $ N = (O, R, W) $, where $O$ represents a set of objects $o_i$ of the network, $R$ represents a set of relations $r_{o_i, o_j}$ between objects, and $W$ represents a set of weights $w_{r_{o_i,o_j}}$ of the relations. We organize the set of network objects into four different types $O = \{O_{B}, O_{T}, O_{S}, O_{M}\}$, where $O_{B}$ are objects that identify each bug report, $O_ {T} $ are terms $t$ extracted from textual data of the bug reports, $O_{S} $ are source files related to bug reports, and $O_{M} $ are code metrics (discretized into intervals) computed from the source code files.

\begin{sloppypar}
The general idea of network regularization is to obtain a new representation $\mathbf{F} \in \mathbb{R}^d$ in the $d$-dimensional vector space model, which satisfies two assumptions: (1) two objects $o_i $ and $o_j $ that share neighbors in the network must have similar vector representation, i.e., $\mathbf{F}(o_i) \sim \mathbf{F}(o_j)$, and (2) term-type objects in the network must have vector representation similar to the word embedding representation, i.e., $\mathbf{F}(t) \sim \mathbf{Y}(t)$.

It is inspired by the theoretical regularization framework of Ji et al.~\cite{Ji2010GRTCOH}. In Equation~\ref{func-reg} we propose a regularization function for the BULNER method, where the goal is to minimize the function according to a representation model $\mathbf{F} $ for all objects of the network, given a word embedding $\mathbf{Y}$.
\end{sloppypar}

\begin{equation}\footnotesize
Q(\mathbf{F}) =
\sum_{{O}_i,{O}_j}^{\{O_{B}, O_{T}, O_{S}, O_{M}\}} \frac{1}{2}
\sum_{o_i\in {O}_i}
\sum_{o_j\in {O}_j} w_{r_{o_i,o_j}}
\big|\mathbf{F}({o_i})-\mathbf{F}({o_j})\big|^2
+\lim_{\mu \rightarrow \infty}\mu\sum_{t\in {O_{T}}}\big|\mathbf{F}(t)-\mathbf{Y}(t)\big|^2
\label{func-reg}
\end{equation}

\begin{sloppypar}
The first term of the regularization function is responsible for the first assumption, in which related objects must have similar representations to minimize the distance $w_{r_{o_i,o_j}} \big|\mathbf{F}({o_i})-\mathbf{F}({o_j})\big|^2$. Regarding the second assumption, the proposed regularization function ensures that the $d$-dimensional representation of a term will remain the same as the word embedding representation, i.e., $\lim_{\mu \rightarrow \infty}\mu\sum_{t\in {O_{T}}}\big|\mathbf{F}(t)-\mathbf{Y}(t)\big|^2$.
\end{sloppypar}

In practical terms, we can minimize the regularization function of the Equation~\ref{func-reg} by using label (information) propagation techniques. In this case, BULNER initializes the representation of the term type objects according to word embedding $\mathbf{Y}$, whereas the representation of the remaining objects of the network can be randomly initialized. In each iteration, BULNER propagates the information of the term objects (i.e., WEmb) to the objects of the bug report type. The information is then propagated from bug reports to source code files objects and then to objects representing code metrics. The information is propagated back and each object $o_i \in O$ adjusts its $\mathbf{F}(o_i)$ representation. This process continues until there are no more significant changes in the $\mathbf{F}$ representation or until it reaches a maximum number of iterations, i.e., until the convergence of the BULNER method in which $\mathbf{F}$ is the learned representation for bug-code vector space model.

After the regularization process, we can directly compute the similarity between a bug report object $o_b \in O_{B}$ and a source code type object $o_s \in O_{S}$ as defined in the cosine similarity of Equation~\ref{burn_cos}.

\begin{equation}
\cos(o_b,o_s) = {\mathbf{F}(o_b) \cdot \mathbf{F}(o_s) \over \|\mathbf{F}(o_b)\| \|\mathbf{F}(o_s)\|}
\label{burn_cos}
\end{equation}

\begin{equation}
sim(r_{new},r_{train}) = (1-\alpha)BOW(r_{new},r_{train}) + \alpha BULNER(r_{new},r_{train})
\label{score1}
\end{equation}

A new bug report can be represented in the bug-code space vector through the word embedding of its terms and thus obtain a representation $\mathbf{F}({o_b})$. Equation~\ref{score1} defines a new similarity function for bug localization in BULNER, which is a linear combination of similarity between bug reports in BoW model and similarity in Bug-Code Vector Space model since new bug reports contain only textual information. While we can define the BoW function as the cosine similarity between bug report keywords, the BULNER function represents the cosine similarity between the bug report representation $\mathbf{F}(o_b)$ and the source code file $\mathbf{F}(o_s)$ (Equation \ref{burn_cos}). The $\alpha$ parameter is a combination factor that allows defining the weight of the bug-code vector space model in the new similarity function, which can be estimated empirically. We use this new similarity function to calculate the scores for source files potentially related to the new bug reports.


\section{Experimental Evaluation}
\label{sec:methodology}

\subsection{Dataset}
\begin{sloppypar}
To evaluate our approach, we obtained data from three open source projects: AspectJ, Birt, and Tomcat. We extracted bug report data associated with each project from the Bugzilla repository provided by Ye~\textit{et~al.}~\cite{Ye2014LTRRFF}, while we mined each project's repository (located in GitHub) to obtain code-related metrics. For each bug report, we checked out a before-fix version of the source code from Github. Then, we used Understand\texttrademark\footnote{Understand\texttrademark: \url{https://scitools.com/}.} to calculate different code metrics (object-oriented, volume and complexity metrics).
\end{sloppypar}


\subsection{Baselines}
\begin{sloppypar}
We consider two methods in the literature for experimental evaluation. The first uses only the BoW model and cosine similarity to retrieve similar bug reports and related source files~\cite{Zhou2012WSTBB}. The second combines the BoW model and WEmb models~\cite{Ye2016FWETDS}.
\end{sloppypar}

\subsection{Evaluation Metrics}
We use Mean Average Precision (MAP) as an evaluation criterion. Equation~\ref{precision} calculates the precision in identifying $NP$ buggy files, given a maximum value of $ k $ recommendations. Equation~\ref{ap} calculates the precision average, where NPI is the total number of positive instances. Equation~\ref{map} calculates the MAP, where M is the total of bug reports. We use MAP with $k = \{1,5,10\}$, presented as MAP@1, MAP@5 and MAP@10.

\noindent \begin{minipage}{.3\linewidth}
\begin{equation}
P(k) = \frac{NB}{k}
\label{precision}
\end{equation}
\end{minipage}%
\begin{minipage}{.3\linewidth}
\begin{equation}
AP = \sum_{i=1}^{N}\frac{P(i)}{NPI}
\label{ap}
\end{equation}
\end{minipage}%
\begin{minipage}{.4\linewidth}
\begin{equation}
MAP = \frac{1}{M}\sum_{j=1}^{M}AP(j)
\label{map}
\end{equation}
\end{minipage}



\section{Results and Discussion}
\label{sec:evaluation}

\subsection{RQ1: How effective is BULNER?}

Table~\ref{tab:highmap} shows the best method's performance (max MAP performance independent of combination factor) for the three datasets. The results suggest that BULNER is the best method for the three datasets.
BULNER achieves this result by combining BoW with WEmb and Network Regularization. It receives as input a heterogeneous network ($ N $), and it treats each type of objects and links separately. Moreover, BULNER minimizes classification error when preserving consistency for each relation graph ($ R $) by applying graph regularization~\cite{Ji2010GRTCOH}. A statistical analysis of the results (Student's t-Tests with 95\% confidence) does not allow us to state that the BULNER method is significantly superior to other methods, mainly due to the few data sets used in the experimental evaluation.



\begin{table}[!ht]
\scriptsize
\caption{MAP Performance Comparison with the State-of-the-art Methods}
\label{tab:highmap}
\begin{tabular}{l|l|l|l|l|l|l|l|l|l|}
\cline{2-10}
                                            & \multicolumn{3}{c|}{AspectJ} & \multicolumn{3}{c|}{Tomcat} & \multicolumn{3}{c|}{Birt} \\ \hline
\multicolumn{1}{|l|}{Methods}               & MAP@1    & MAP@5   & MAP@10  & MAP@1   & MAP@5   & MAP@10  & MAP@1   & MAP@5  & MAP@10 \\ \hline
\multicolumn{1}{|l|}{BoW+Cosine} & 0.1185   & 0.1738  & 0.1879  & 0.2121  & 0.2908  & 0.3017  & 0.0900  & 0.1372 & 0.1477 \\ \hline
\multicolumn{1}{|l|}{Embedding}             & 0.1185   & 0.1738  & 0.1811  & 0.2134  & 0.2908  & 0.3001  & 0.0928  & 0.1402 & 0.1504 \\ \hline
\multicolumn{1}{|l|}{Bulner}                & 0.1390   & 0.1913  & 0.2059  & 0.2201  & 0.2952  & 0.3078  & 0.0968  & 0.1420 & 0.1525 \\ \hline
\end{tabular}
\end{table}

\subsection{RQ2: What is the contribution of each method?}
\begin{figure*}[!hbt]
\centering
\subfigure[\label{map1_aspectj}{AspectJ (MAP@1)}]
{\includegraphics[width=0.32\textwidth]{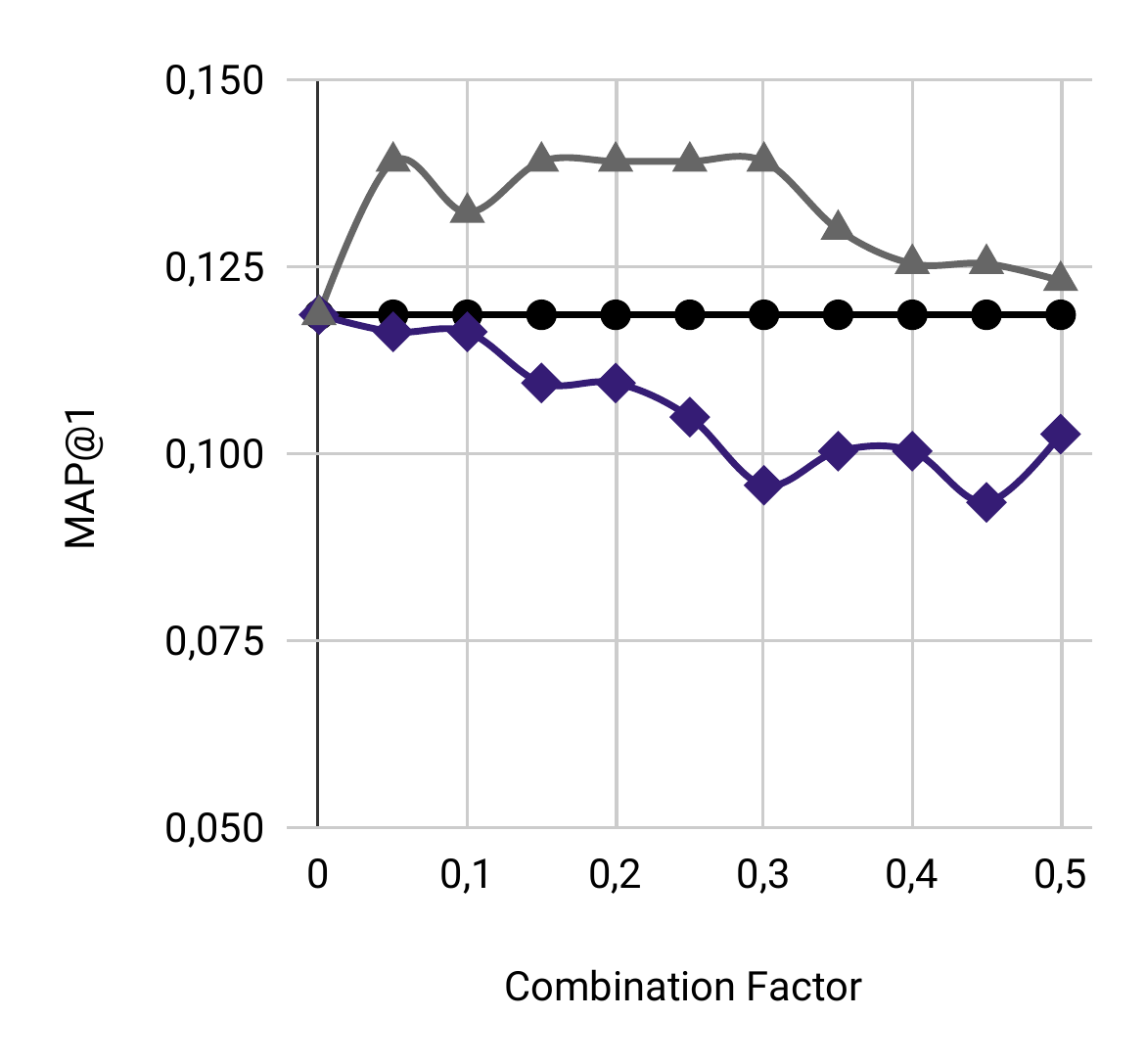}}
	\vspace{1pt}
\subfigure[\label{map1_birt}{Birt (MAP@1)}]
{\includegraphics[width=0.32\textwidth]{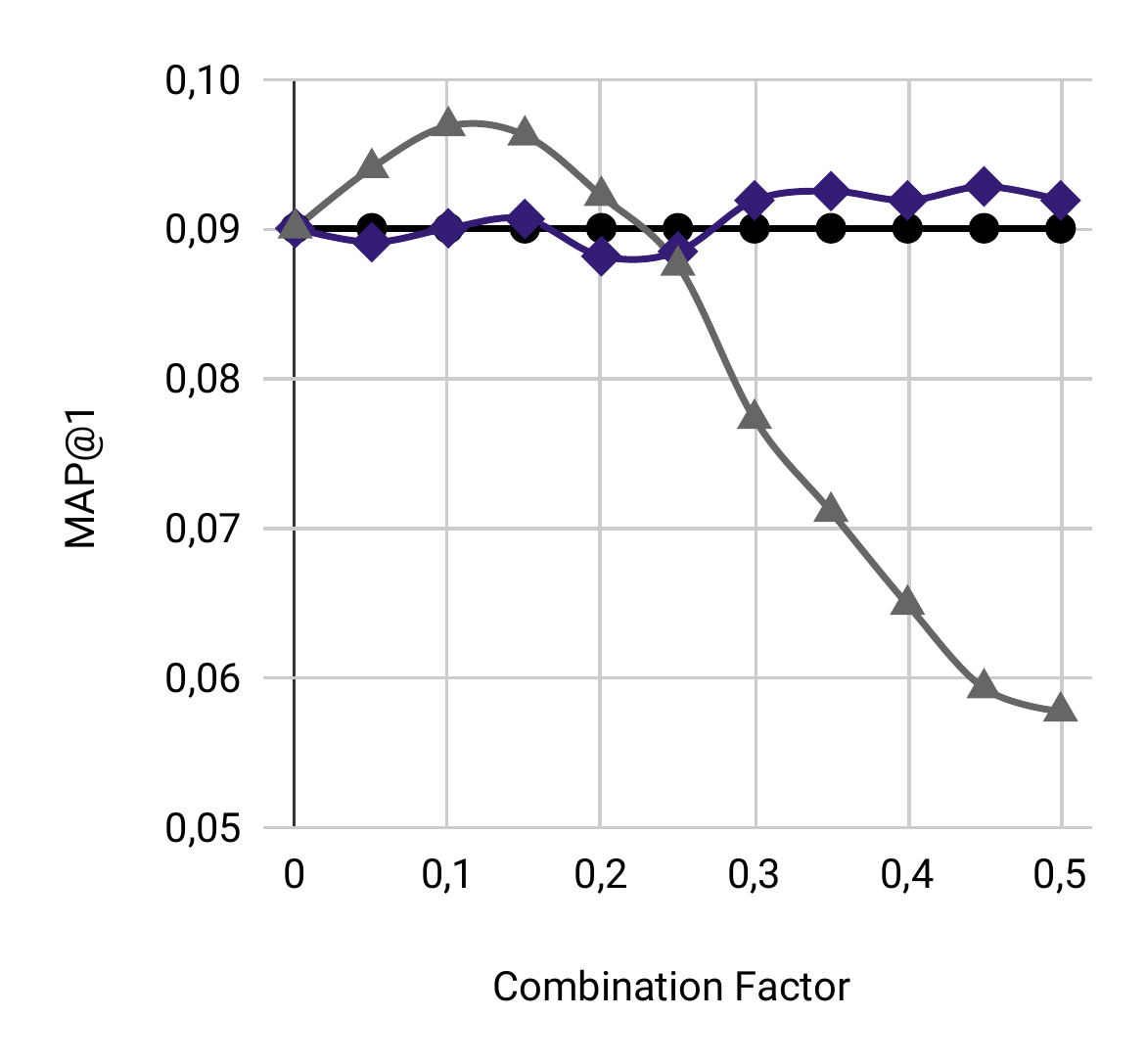}}
	\vspace{1pt}
\subfigure[\label{map1_tomcat}{Tomcat (MAP@1)}]
{\includegraphics[width=0.32\textwidth]{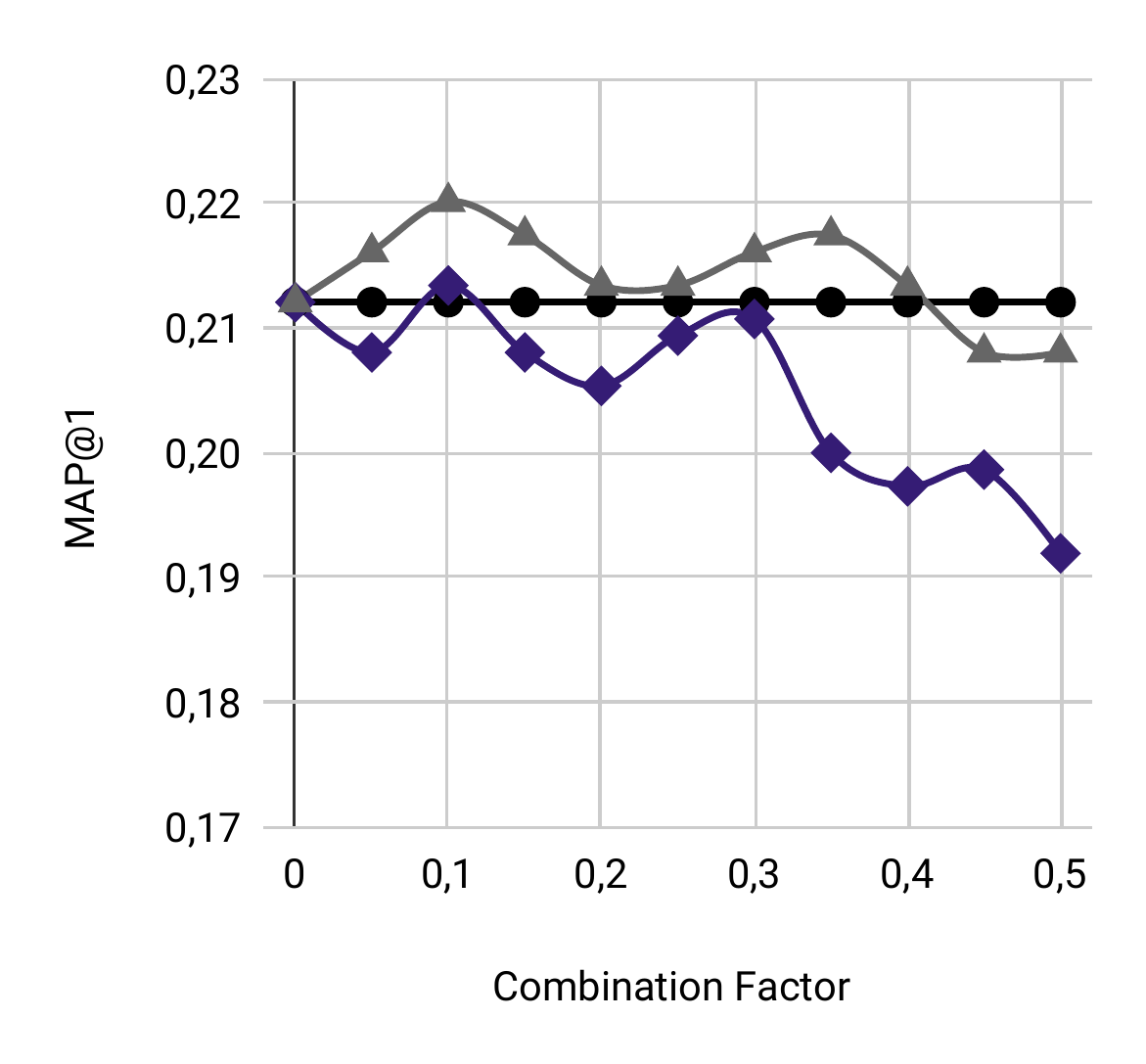}}
	\vspace{1pt}
\subfigure[\label{map5_aspectj}{AspectJ (MAP@5)}]
{\includegraphics[width=0.32\textwidth]{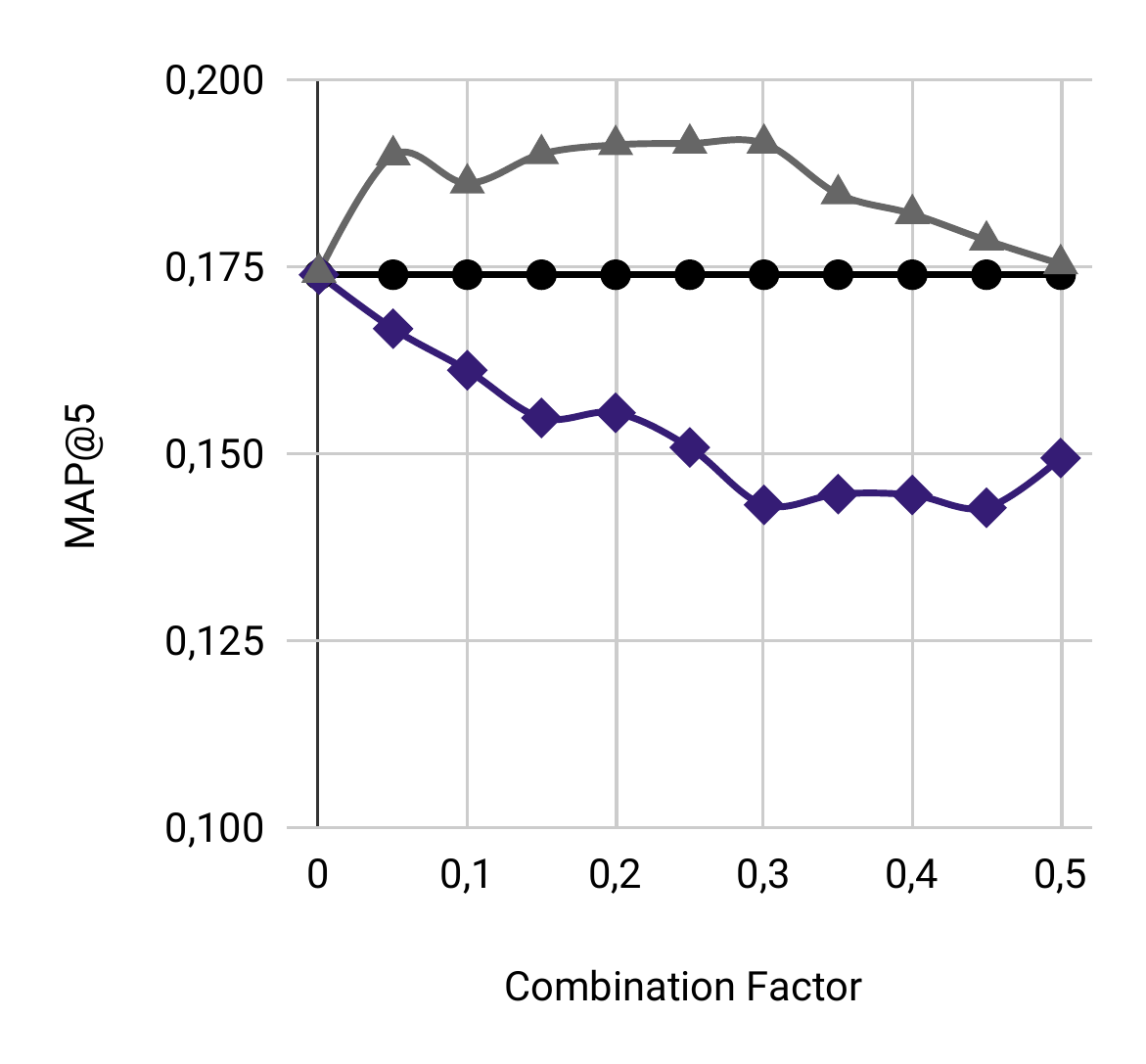}}
	\vspace{1pt}
\subfigure[\label{map5_birt}{Birt (MAP@5)}]
{\includegraphics[width=0.32\textwidth]{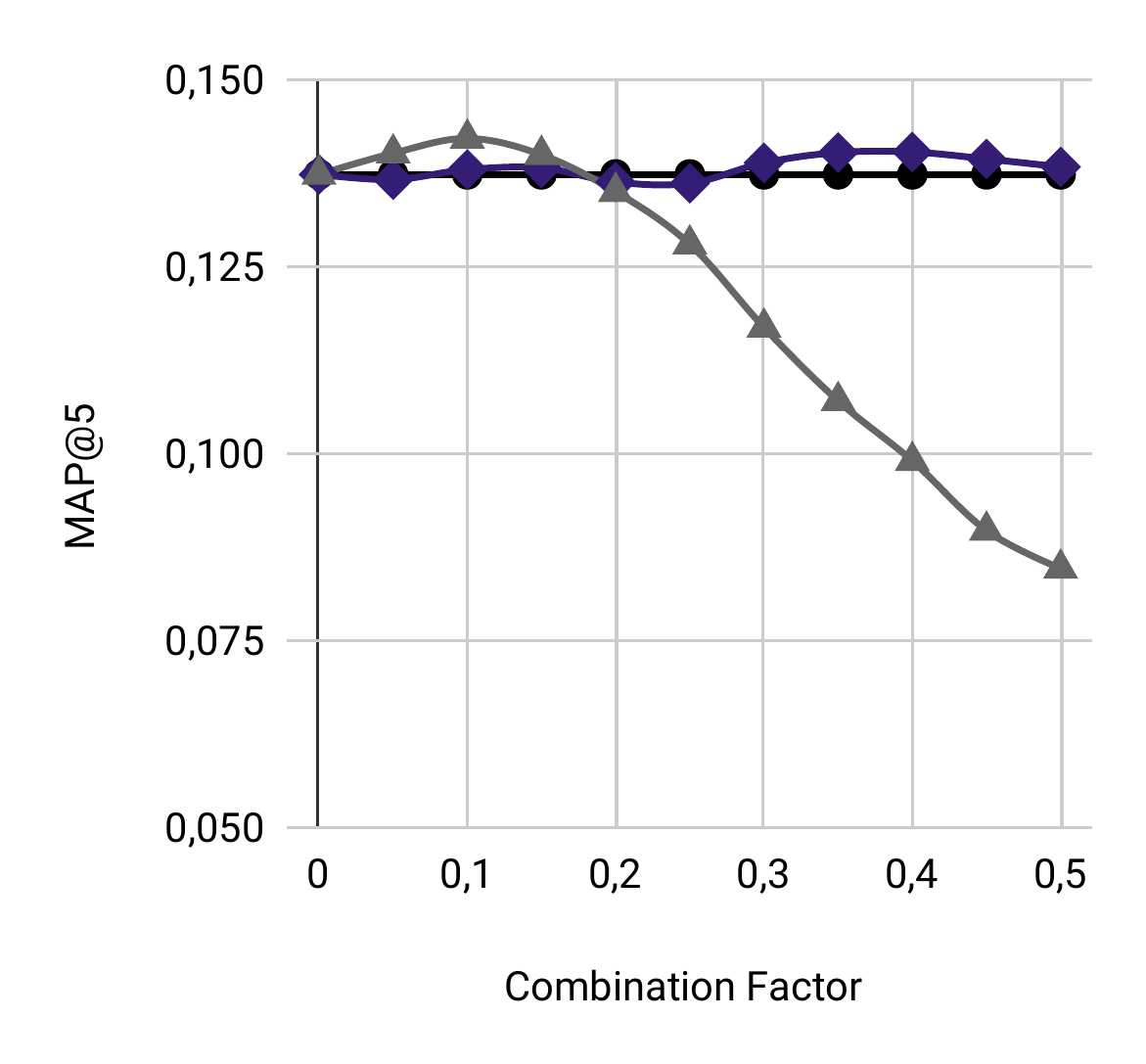}}
	\vspace{1pt}
\subfigure[\label{map5_tomcat}{Tomcat (MAP@5)}]
{\includegraphics[width=0.32\textwidth]{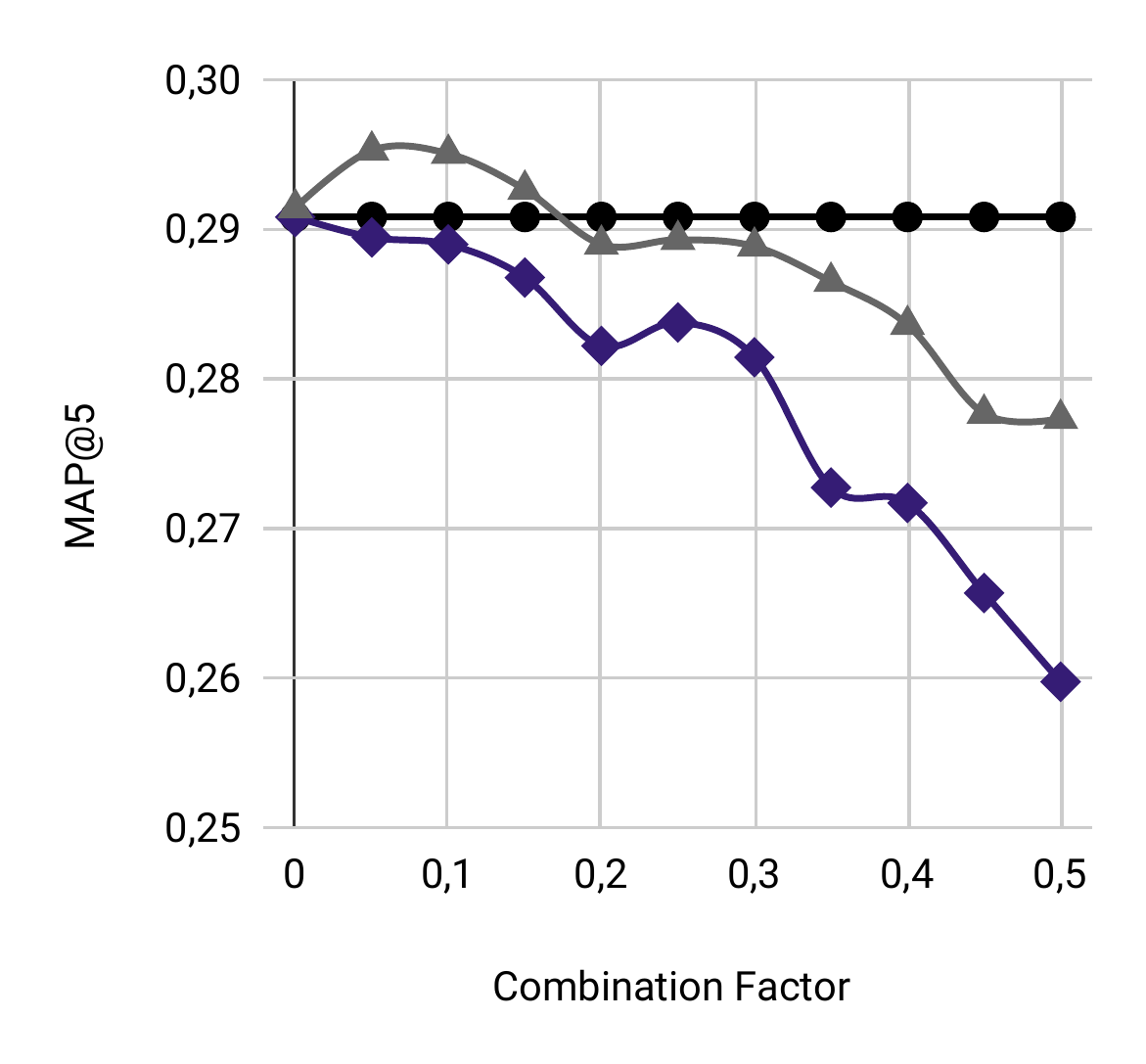}}
	\vspace{1pt}
\subfigure[\label{map10_aspectj}{AspectJ (MAP@10)}]
{\includegraphics[width=0.32\textwidth]{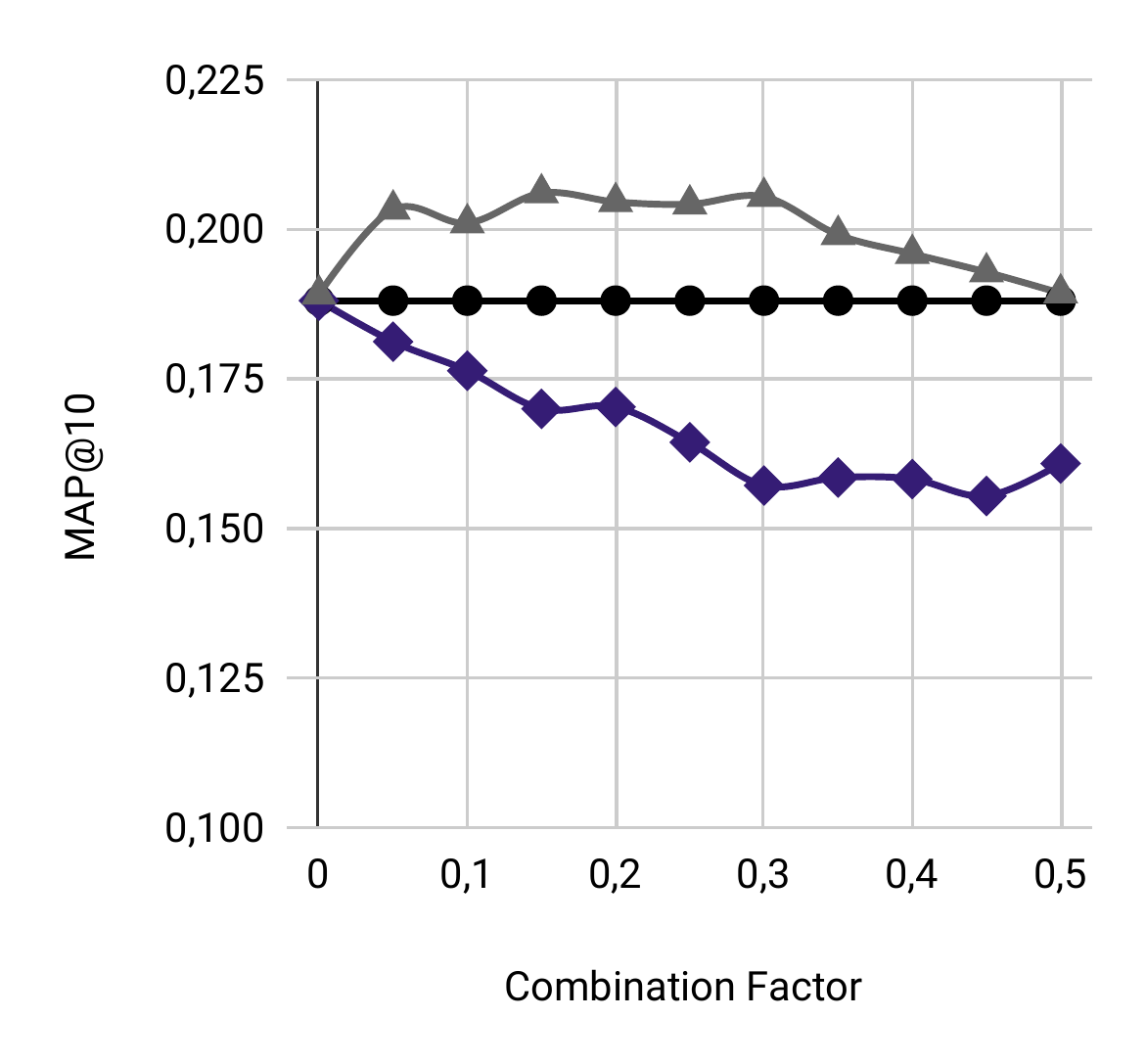}}
	\vspace{1pt}
\subfigure[\label{map10_birt}{Birt (MAP@10)}]
{\includegraphics[width=0.32\textwidth]{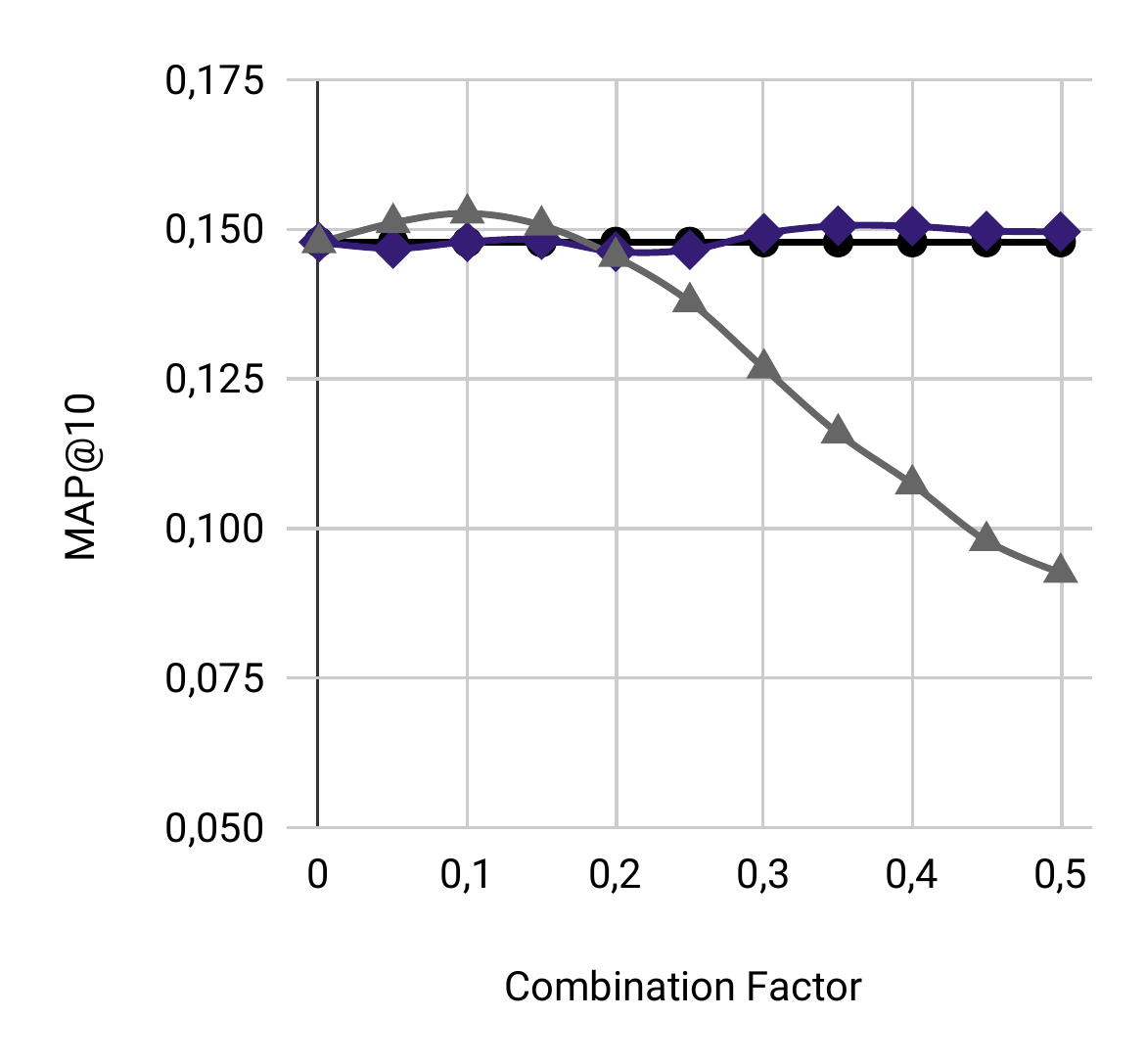}}
	\vspace{1pt}
\subfigure[\label{map10_tomcat}{Tomcat (MAP@10)}]
{\includegraphics[width=0.32\textwidth]{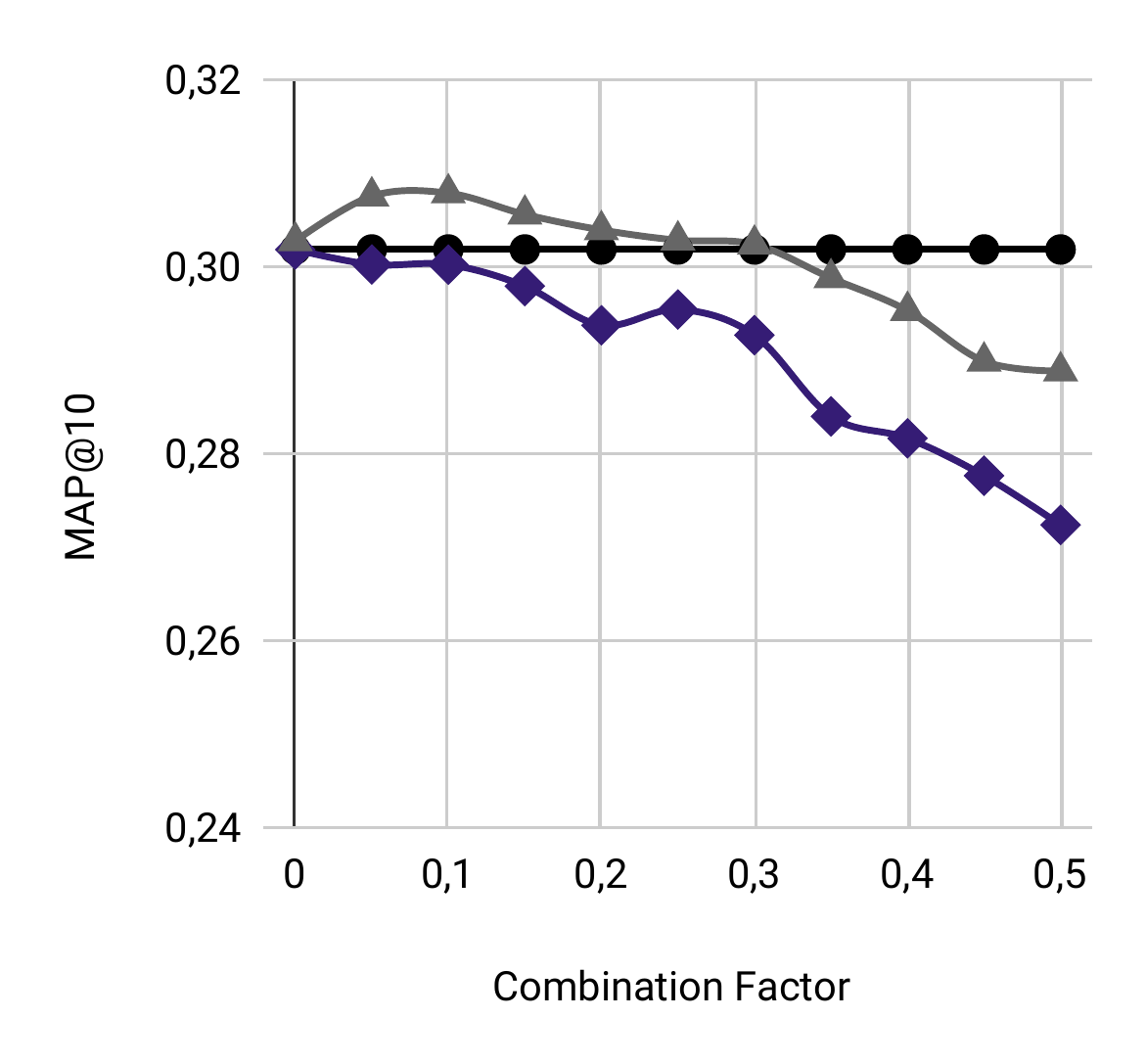}}
\caption{Methods' performance. $\medblacktriangleup$ BULNER; $\mdlgblkcircle$~BoW+Cosine; $\medblackdiamond$~Embedding.}
\label{map2}
\end{figure*}

We evaluate the contributions of each method by the combination factor ($\alpha$). In Figure~\ref{map2}, when $\alpha$=0, all methods have the performance equal to baseline (BoW+Cosine), but when we increment $\alpha$, each method has different behavior. In general, BULNER has better performance for: AspectJ when  0.15 < $\alpha$ < 0.3; Birt when $\alpha$ = 0.1 and Tomcat when 0.05 < $\alpha$ < 0.1. These variations between software project occur because each one has its context.  


\subsection{Threats to Validity}
\label{sec:threats}
Regarding \textbf{Internal validity}, as only bug reports with ``resolved'' status were selected (because they represent bug reports that span the entire life cycle), the proposed model did not consider an immature bug report, such as those newly created by stakeholders. This restriction may have an impact on the performance of the proposed model. Regarding \textbf{External validity}, the results of this study can not be generalized to proprietary software since only Open Source software projects were analyzed. Finally, concerning \textbf{Conclusion validity}, we choose dataset provided by Ye~\textit{et~al.} that is largely used to maximize the quality of the data collected.

\section{Related Work}
\label{sec:related}
\begin{sloppypar}
Zhou~\textit{et~al}. propose an IR-based bug localization tool that using a revised Vector Space Model (rVSM) to represent software documents (bug report and source code)~\cite{Zhou2012WSTBB}. In our study, we define a baseline using the Vector Space Model to representing data sources.
\end{sloppypar}

\begin{sloppypar}
Wen~\textit{et~al.} define an IR-based bug localization model that combining three models: the natural language model, code entity names model, and Boosting model~\cite{Wen2016LLBFSC}. In our work, we also evaluate the impact of combining models. Firstly, we only used the embedding model in BULNER, and then we incremental added value of regularization and combined it with the embedding model.
\end{sloppypar}

\vspace{1pt}
\section{Conclusion}
\label{sec:conclusion}
We proposed a method BULNER, which locates bugs in terms of source files from bug reports and source code data. Our method is competitive with two other state-of-the-art methods. BULNER is very promising. In one case, it can recommend the 30\% suspicious file within top 5 for one bug report. For future works, we intend to compare our work with different types of network embedding methods, for example, network embedding with side information or advanced information preserving network embedding. Additionally, we plan to extend our dataset with source code change genealogy and evaluate the impact on the performance of the methods. Our BULNER source code, as well as the datasets used, are publicly available at https://github.com/jacsonrbinf/bulner.

\bibliographystyle{sbc}
\bibliography{sbc-template}

\end{document}